\documentclass[aps,prc,preprint,showpacs,showkeys]{revtex4}
\usepackage{amssymb}
\usepackage{amsmath}
\usepackage{graphicx}
\usepackage{bm}
\usepackage{epsf}

\begin{document}

\title{Superfluidity of $\Lambda$ hyperons in neutron stars}
\author{Y. N. Wang}
\affiliation{Department of Physics, Nankai University, Tianjin 300071,
People's Republic of China}
\author{H. Shen}
\email{songtc@nankai.edu.cn}
\affiliation{Department of Physics, Nankai University, Tianjin 300071,
People's Republic of China}

\begin{abstract}
We study the $^1S_0$ superfluidity of $\Lambda$ hyperons in neutron star
matter and neutron stars. We use the relativistic mean field (RMF)
theory to calculate the properties of neutron star matter.
In the RMF approach, the meson-hyperon couplings are constrained by
reasonable hyperon potentials that include the updated information from
recent developments in hypernuclear physics.
To examine the $^1S_0$ pairing gap of $\Lambda$ hyperons,
we employ several $\Lambda\Lambda$ interactions based on
the Nijmegen models and used in double-$\Lambda$ hypernuclei studies.
It is found that the maximal pairing gap obtained is a few tenths
of a MeV.
The magnitude and the density region of the pairing gap
are dependent on the $\Lambda\Lambda$ interaction and the treatment
of neutron star matter. We calculate neutron star properties
and find that whether the $^1S_0 $ superfluidity of $\Lambda$ hyperons
exists in the core of neutron stars mainly depends on the $\Lambda\Lambda$
interaction used.
\end{abstract}

\maketitle



\section{Introduction}

\label{sec:1}

Neutron stars are natural laboratories for studying the physics of dense
matter. They are associated with some of the most exotic environments in the
universe. Our knowledge of neutron star interiors is still uncertain.
The central density of neutron stars can be extremely high, and many
possibilities for such dense matter have been suggested~\cite{Weber05,PR00,PR07}.
For densities below twice normal nuclear matter density
($\rho_{0} \sim 0.15\;\textrm{fm}^{-3}$), the matter consists of
only nucleons and leptons. When the density is higher than $2 \rho_{0}$,
the equation of state (EOS) and composition of matter are much less certain.
The presence of hyperons in neutron stars has been studied by many
authors~\cite{PRC96,PRC99,Panda04,Shen02,Gl01}. $K^{-}$ condensation in dense
matter was suggested by Kaplan and Nelson~\cite{KN86} and has been extensively discussed
in many works~\cite{PR00,PRC96,YS08}. It has been suggested that the quark
matter may exist in the core of massive neutron stars, and the hadron-quark
phase transition can proceed through a mixed phase of hadronic and quark
matter~\cite{Weber05,PR00,Panda04,PRC08}. If deconfined quark matter does
exist inside stars, it is likely to be in a color superconducting
phase~\cite{Weber05,Panda04}, and various color superconducting phases
have been intensively investigated in recent years~\cite{PR05,Huang,Zhuang}.

Baryon pairing is believed to play an important role in the evolution of
neutron stars~\cite{PR00,RMP03,PRL00}. The presence of neutron superfluidity in
the crust and the inner part of neutron stars can be considered well
established~\cite{RMP03}. The neutron fluid in the crust probably forms
a $^1S_0$ superfluid. With increasing density, the $^1S_0$ interaction turns
repulsive, and the neutrons in the outer core mainly form a $^3P_2$
superfluid. On the other hand, one can expect $^1S_0$ proton pairing in the
outer core because the small proton fraction brings about a low proton
density in this region. In the inner core of neutron stars, hyperons may
appear through the weak interaction because of the fast rise of the baryon
chemical potentials with density. It is widely accepted that hyperons
appear around $2 \rho_{0}$. The presence of hyperons tends to soften
the EOS at high density and lower the maximum mass of neutron
stars~\cite{PRC96,PRC99,Panda04,Shen02,Gl01}
as well as increase the neutron star cooling rate~\cite{PRL00,APJS99}.
In general, the first hyperon to appear is $\Lambda$,
which is the lightest one with an attractive potential in nuclear matter.
The potential of $\Sigma$ hyperons
is now considered to be repulsive; therefore $\Sigma^-$ appears at a higher
density than $\Lambda$ in neutron star matter~\cite{JPG08,YS09}.
The $^1S_0$ superfluidity of $\Lambda$ hyperons is suggested to occur in
the same way as that of neutrons arising from the attractive $\Lambda\Lambda$
interaction in the $^1S_0$ channel~\cite{BB98,TT99,TT00,TMS03,TNYT06}.
It is known that hyperon pairing can significantly affect the thermal
evolution of neutron stars by suppressing neutrino emission from the
hyperon direct Urca process~\cite{TNYT06,APJ98,APJ09}.
Young neutron stars cool primarily by neutrino emission from the interior.
As discussed in Refs.~\cite{Page06,Yakovlev04,Yakovlev01},
the neutrino emissivity in superfluid matter is exponentially suppressed
when the temperature $T$ is much lower than the superfluid critical
temperature $T_c$. On the other hand, superfluidity initiates a specific
neutrino emission from the Cooper pair breaking and formation process,
which is forbidden in nonsuperfluid matter.
This process is exponentially suppressed when $T\ll T_c$, and it is much
less efficient than the direct Urca process~\cite{Yakovlev01,PLB06}.
Hence the presence of baryon superfluidity can
drastically suppress the neutrino emission, which may play a key role in
neutron star cooling. We are mainly interested in the possibility of $^1S_0$
superfluidity of $\Lambda$ hyperons in neutron stars.
So far, the $^1S_0$ pairing gap of $\Lambda$ hyperons is still uncertain
because it can be significantly influenced by both the properties of matter
and the $\Lambda\Lambda$ interaction. More studies are needed to determine
these uncertain factors using available information from recent
developments in hypernuclear physics.

In this article, we focus on the $^1S_0$ superfluidity of $\Lambda$ hyperons
in neutron star matter, which is composed of a chemically equilibrated and
charge-neutral mixture of nucleons, hyperons, and leptons. To calculate the
pairing gap, we need to specify how to treat the neutron star matter and the
$\Lambda\Lambda$ interaction. In this article, we use the relativistic
mean field (RMF) theory to calculate the properties of neutron star matter.
The RMF theory has been successfully and widely used for the description of
nuclear matter and finite nuclei~\cite{Serot86,Ring90,Ren02,Toki96,Shen06}.
It has also been applied to providing the EOS of dense matter for use in
supernovae and neutron stars~\cite{Shen98}. In the RMF approach, baryons
interact through the exchange of scalar and vector mesons. The meson-nucleon
coupling constants are generally determined by fitting to some nuclear
matter properties or ground-state properties of finite nuclei.
To examine the influence of the RMF parameters, we employ two
successful parameter sets, TM1~\cite{ST94} and NL3~\cite{NL3},
which have been widely used in many studies of nuclear
physics~\cite{PRC08,Shen98,Shen06,ST94,NL3,HTT95}.
As for the meson-hyperon couplings, there are large uncertainties because of
limited experimental data in hypernuclear physics. Generally, one can use the
coupling constants derived from the quark model or the values constrained by
reasonable hyperon potentials. The meson-hyperon couplings play an important
role in determining the properties of neutron star matter~\cite{Gl01,Shen03}.
We use the values constrained by reasonable hyperon potentials that
include the updated information from recent developments in hypernuclear
physics. We take into account the two additional hidden-strangeness
mesons, $\sigma^{\ast}$ and $\phi$, which were originally introduced to
obtain the strong attractive $\Lambda\Lambda$ interaction deduced from the
earlier measurement~\cite{PRL93}. A recent observation of the double-$\Lambda$
hypernucleus $_{\Lambda\Lambda}^{6}\textrm{He}$, called the Nagara
event~\cite{Nagara}, has had a significant impact on strangeness nuclear physics.
The Nagara event provides unambiguous identification of
$_{\Lambda\Lambda}^{6}\textrm{He}$ production with a precise $\Lambda\Lambda$
binding energy value
$B_{\Lambda\Lambda }=7.25\pm 0.19_{-0.11}^{+0.18}\;\textrm{MeV}$, which suggests
that the effective $\Lambda\Lambda$ interaction should be considerably
weaker ($\triangle B_{\Lambda\Lambda}\simeq 1\;\textrm{MeV}$) than that
deduced from the earlier measurement
($\triangle B_{\Lambda\Lambda}\simeq$ 4--5 MeV).
The weak hyperon-hyperon ($YY$) interaction suggested by the Nagara event
has been used to reinvestigate the properties of multistrange systems,
and it has been found that the change of $YY$ interactions affects
the properties of strange hadronic matter
dramatically~\cite{PRC08,PRC03s,JPG04s,JPG05}.
We would like to examine whether the $^1S_0 $ superfluidity of $\Lambda$
hyperons exists in neutron star matter, and how large the pairing gap
can be if it does, by considering recent developments in hypernuclear physics.

The aim of this article is to investigate the possibility of forming
a $\Lambda$ superfluid in neutron stars. It has been suggested
that hyperon superfluidity could significantly suppress the neutrino emission
in the core of a neutron star and play a key role in neutron star
cooling~\cite{TNYT06,APJ98}.
Over the last decade, there has been some discussion in the literature
about hyperon pairing in dense matter~\cite{BB98,TT99,TT00,TMS03,TNYT06,SIGMA04}.
In the work of Balberg and Barnea~\cite{BB98}, the $^1S_0$ superfluidity of $\Lambda$
hyperons has been studied by using an effective $\Lambda\Lambda$ interaction
based on a $G$ matrix calculation and an approximation of
nonrelativistic effective mass obtained from single-particle energies.
Their calculation predicts a gap energy of a few tenths of a MeV
for $\Lambda$ Fermi momenta up to about $1.3$ $\textrm{fm}^{-1}$.
In Refs.~\cite{TT99,TT00}, Takatsuka and Tamagaki studied $\Lambda$ superfluidity
using two types of bare $\Lambda\Lambda$ interactions based on the
one-boson-exchange (OBE) model and two types of hyperon core models.
They found that $\Lambda$ superfluidity could exist in a density region
between 2$\rho_{0}$ and (2.6--4.6)$\rho_{0}$, depending on the pairing
interaction and the hyperon core model.
A study of $\Lambda\Lambda$ pairing in a pure neutron background
has been presented by Tanigawa \emph{et al.}~\cite{TMS03} using the
relativistic Hartree-Bogoliubov model, where the $\Lambda\Lambda$ pairing gap
was found to decrease with increasing background density
and decreasing $\Lambda\Lambda$ attraction.
Both $\Lambda$ and $\Sigma^-$ superfluidities in neutron star matter have been
investigated by Takatsuka \emph{et al.}~\cite{TNYT06} using three pairing
interactions based on the OBE model and several nonrelativistic EOS with different
incompressibilities. It was found that both $\Lambda$ and $\Sigma^-$ are
superfluid as soon as they begin to appear at around $4 \rho_{0}$,
although the pairing gap and the density region depend on the pairing
interaction and the EOS of neutron star matter.
The effect of the Nagara event on $\Lambda$ superfluidity has been discussed
in Refs.~\cite{TMS03,TNYT06}, where the weak attractive $\Lambda\Lambda$
interaction suggested by the Nagara event leads to very small pairing
gap in dense neutron matter~\cite{TMS03} or the disappearance of
$\Lambda$ superfluidity in neutron star matter~\cite{TNYT06}.
All these studies indicate that the $\Lambda\Lambda$ pairing gap
in dense matter depends both on the $\Lambda\Lambda$ interaction
and on the EOS of matter. The key role of $\Lambda$ superfluidity
in neutron star cooling motivates us to investigate the possibility of
forming a $\Lambda$ superfluid in neutron stars by carefully considering
the pairing interaction and the description of neutron star matter
with the updated information from recent developments in
hypernuclear physics.

This article is arranged as follows. In Sec.~\ref{sec:2}, we briefly
describe the RMF theory for the calculation of neutron star matter
properties. In Sec.~\ref{sec:3}, we discuss the $\Lambda\Lambda$
interaction used in the gap equation. We present the numerical results
in Sec.~\ref{sec:4}. Section~\ref{sec:5} is devoted to a summary.


\section{ Relativistic mean field theory}

\label{sec:2}

We use the RMF theory to describe the neutron star matter, which is composed
of a chemically equilibrated and charge-neutral mixture of nucleons,
hyperons, and leptons. In the RMF approach, baryons interact through the
exchange of scalar and vector mesons. The baryons considered in this work
are nucleons ($p$ and $n$) and hyperons ($\Lambda$, $\Sigma$, and $\Xi$).
The exchanged mesons include isoscalar scalar and vector mesons
($\sigma$ and $\omega$), an isovector vector meson ($\rho$), and two additional
hidden-strangeness mesons ($\sigma^{\ast }$ and $\phi$).
The total Lagrangian density of neutron star matter takes the form
\begin{eqnarray}
\mathcal{L}_{RMF} &=&\sum_{B}\bar{\psi}_{B}\left[ i\gamma_{\mu}
\partial^{\mu}-m_{B}-g_{\sigma B}\sigma-g_{\sigma^{\ast}B} \sigma^{\ast}
-g_{\omega B}\gamma_{\mu}\omega^{\mu}\right.  \notag \\
&&\left. -g_{\phi B}\gamma_{\mu}\phi^{\mu}-g_{\rho B}\gamma_{\mu}
\tau_{iB}\rho_{i}^{\mu}\right] \psi_{B} +\frac{1}{2}\partial_{\mu}
\sigma\partial^{\mu}\sigma -\frac{1}{2}m_{\sigma}^{2}\sigma^{2}  \notag \\
&&-\frac{1}{3}g_{2}\sigma^{3}-\frac{1}{4}g_{3}\sigma^{4}
-\frac{1}{4}W_{\mu\nu}W^{\mu\nu}+\frac{1}{2}m_{\omega}^{2}
\omega_{\mu}\omega^{\mu} \notag \\
&&+\frac{1}{4}c_{3}\left( \omega_{\mu}\omega^{\mu}\right)^{2}
-\frac{1}{4}R_{i\mu\nu}R_{i}^{\mu\nu}+\frac{1}{2}m_{\rho}^{2}
\rho_{i\mu}\rho_{i}^{\mu} \notag \\
&&+\frac{1}{2}\partial_{\mu}\sigma^{\ast}\partial^{\mu}\sigma^{\ast}
-\frac{1}{2}m_{\sigma^{\ast}}^{2}\sigma^{\ast 2}-\frac{1}{4} S_{\mu\nu}S^{\mu\nu}
+\frac{1}{2}m_{\phi}^{2}\phi_{\mu}\phi^{\mu}  \notag \\
&&+\sum_{l}\bar{\psi}_{l}\left[ i\gamma_{\mu}\partial^{\mu}-m_{l}\right]
\psi_{l},
\label{eq:Lrmf}
\end{eqnarray}
where $\psi_{B}$ and $\psi_{l}$ are the baryon and lepton fields,
respectively. The index $B$ runs over the baryon octet
($p$, $n$, $\Lambda$, $\Sigma^{+}$, $\Sigma^{0}$, $\Sigma^{-}$,
$\Xi^{0}$, $\Xi^{-}$), and the sum on $l$ is over electrons and muons
($e^{-}$ and $\mu^{-}$). The field tensors of the vector mesons,
$\omega$, $\rho$, and $\phi$, are denoted by $W_{\mu\nu}$, $R_{i\mu\nu}$,
and $S_{\mu\nu}$, respectively. In the RMF approach, the meson fields are
treated as classical fields, and the field operators are replaced by their
expectation values. The meson field equations in uniform matter have the
following form:
\begin{eqnarray}
&&m_{\sigma }^{2}\sigma +g_{2}\sigma ^{2}+g_{3}\sigma ^{3}
=-\sum_{B}\frac{g_{\sigma B}}{\pi ^{2}}\int_{0}^{k_{F}^{B}}
\frac{m_{B}^{\ast }}{\sqrt{k^{2}+m_{B}^{\ast 2}}}k^{2}dk,
\label{eq:s} \\
&&m_{\omega }^{2}\omega +c_{3}\omega ^{3}=\sum_{B}\frac{g_{\omega B}
\left(k_{F}^{B}\right)^{3}}{3\pi ^{2}},
\label{eq:w} \\
&&m_{\rho }^{2}\rho =\sum_{B}\frac{g_{\rho B}\tau _{3B}
\left(k_{F}^{B}\right) ^{3}}{3\pi ^{2}},
\label{eq:r} \\
&&m_{\sigma ^{\ast }}^{2}\sigma ^{\ast }=-\sum_{B}
\frac{g_{\sigma ^{\ast }B}}{\pi ^{2}}\int_{0}^{k_{F}^{B}}
\frac{m_{B}^{\ast }}{\sqrt{k^{2}+m_{B}^{\ast 2}}}k^{2}dk,
\label{eq:ss} \\
&&m_{\phi }^{2}\phi =\sum_{B}\frac{g_{\phi B}\left( k_{F}^{B}\right)^{3}}
{3\pi ^{2}},
\label{eq:ws}
\end{eqnarray}
where $\sigma =\left\langle \sigma \right\rangle$,
      $\omega =\left\langle \omega^{0}\right\rangle$,
      $\rho =\left\langle \rho_{30}\right\rangle$,
      $\sigma^{\ast} =\left\langle \sigma^{\ast}\right\rangle$,
  and $\phi=\left\langle \phi^{0}\right\rangle$
are the nonvanishing expectation values of meson fields in uniform matter;
$m_{B}^{\ast}=m_{B}+g_{\sigma B}\sigma+g_{\sigma^{\ast}B}\sigma^{\ast}$
is the effective mass of the baryon species $B$, and $k_{F}^{B}$ is the
corresponding Fermi momentum.

The meson-baryon coupling constants play an important role in determining
the properties of neutron star matter.
To examine the influence of the RMF parameters, we employ two
successful parameter sets, TM1~\cite{ST94} and NL3~\cite{NL3},
in the present calculation.
These parameters have been determined by fitting to some ground-state
properties of finite nuclei, and they can provide a good description of
nuclear matter and finite nuclei, including unstable nuclei.
With the TM1 (NL3) parameter set, the nuclear matter saturation
density is $0.145$ fm$^{-3}$ ($0.148$ fm$^{-3}$),
the energy per nucleon is $-16.3$ MeV ($-16.3$ MeV),
the symmetry energy is $36.9$ MeV ($37.4$ MeV),
and the incompressibility is $281$ MeV ($272$ MeV)~\cite{ST94,NL3}.
As for the meson-hyperon couplings, we take the
naive quark model values for the vector coupling constants;
\begin{eqnarray}
&&\frac{1}{3}g_{\omega N}=\frac{1}{2}g_{\omega \Lambda }=\frac{1}{2}
g_{\omega \Sigma }=g_{\omega \Xi },  \notag \\
&&g_{\rho N}=\frac{1}{2}g_{\rho \Sigma }=g_{\rho \Xi },
\ \ g_{\rho\Lambda}=0,  \notag \\
&&2g_{\phi \Lambda }=2g_{\phi \Sigma }=g_{\phi \Xi }=-\frac{2\sqrt{2}}{3}
g_{\omega N},\ \ g_{\phi N}=0.
\end{eqnarray}
The scalar coupling constants are chosen to give reasonable hyperon
potentials. We denote the potential depth of the hyperon species $i$ in
the matter of the baryon species $j$ by $U_{i}^{\left( j\right) }$.
It is estimated from the experimental data of single-$\Lambda$ hypernuclei
that the potential depth of a $\Lambda$ in saturated nuclear matter should
be around $U_{\Lambda}^{\left(N\right)} \simeq -30$ MeV~\cite{PRC00}.
For $\Sigma$ hyperons, the analysis of $\Sigma$ atomic experimental data
suggests that $\Sigma$-nucleus potentials have a repulsion inside the nuclear
surface and an attraction outside the nucleus with a sizable absorption.
In recent theoretical works, the $\Sigma$ potential in saturated nuclear
matter is considered to be repulsive with a strength of about
$30$ MeV~\cite{JPG08,PRC00}. Some recent developments in hypernuclear physics
suggest that $\Xi$ hyperons in saturated nuclear matter have an attraction
of around $15$ MeV~\cite{JPG08,PRC00xi}. In this article, we adopt
$U_{\Lambda }^{\left(N\right)}=-30$ MeV,
$U_{\Sigma }^{\left(N\right) }=+30$ MeV, and
$U_{\Xi}^{\left(N\right)}=-15$ MeV
to determine the scalar coupling constants.
We obtain, for the TM1 (NL3) parameter set,
$g_{\sigma\Lambda}=6.228$ (6.323),
$g_{\sigma\Sigma}=4.472$ (4.709), and
$g_{\sigma\Xi}=3.114$ (3.161), respectively.
The hyperon couplings to the hidden-strangeness meson
$\sigma^{\ast}$ are restricted by the relation
$U_{\Xi }^{\left(\Xi\right)} \approx
U_{\Lambda}^{\left(\Xi\right) } \approx
2U_{\Xi}^{\left(\Lambda\right) } \approx
2U_{\Lambda}^{\left(\Lambda\right) }$ obtained in Ref.~\cite{ANN94}.
The weak $YY$ interaction implied by the Nagara event suggests
$U_{\Lambda}^{(\Lambda)}\simeq -5$ MeV, and hence we obtain
$g_{\sigma^*\Lambda}=5.499$ (5.678)
and $g_{\sigma^*\Xi}=11.655$ (11.899) for the TM1 (NL3) parameter set.
We assume $g_{\sigma^*\Sigma}=g_{\sigma^*\Lambda}$
and take $m_{\sigma^{\ast }}=980$ MeV and $m_{\phi }=1020$
MeV in this article.

For neutron star matter consisting of a neutral mixture of baryons and
leptons, the $\beta$-equilibrium conditions without trapped neutrinos
are given by
\begin{eqnarray}
&&\mu_{p}=\mu_{\Sigma^{+}}=\mu_{n}-\mu_{e},  \label{eq:beta1} \\
&&\mu_{\Lambda}=\mu_{\Sigma^{0}}=\mu_{\Xi^{0}}=\mu_{n},  \label{eq:beta2} \\
&&\mu_{\Sigma^{-}}=\mu_{\Xi^{-}}=\mu_{n}+\mu_{e},  \label{eq:beta3} \\
&&\mu_{\mu}=\mu_{e},  \label{eq:beta4}
\end{eqnarray}
where $\mu_{i}$ is the chemical potential of species $i$.
At zero temperature, the chemical potentials of baryons and leptons are
given by
\begin{eqnarray}
&&\mu_{B}=\sqrt{{k_{F}^{B}}^{2}+m_{B}^{\ast 2}}+g_{\omega B}\omega
+g_{\phi B}\phi +g_{\rho B}\tau_{3B}\rho ,  \label{eq:mub} \\
&&\mu_{l}=\sqrt{{k_{F}^{l}}^{2}+m_{l}^{2}},  \label{eq:mul}
\end{eqnarray}
respectively. The electric charge neutrality condition is expressed by
\begin{eqnarray}
\rho_{p}+\rho_{\Sigma^{+}} =\rho_{e}+\rho_{\mu}
+\rho_{\Sigma^{-}}+\rho_{\Xi^{-}},  \label{eq:charge}
\end{eqnarray}
where $\rho_{i}=\left( k_{F}^{i}\right)^{3}/(3\pi^{2})$ is the number
density of species $i$. We solve the coupled
Eqs.~(\ref{eq:s})--(\ref{eq:ws}), (\ref{eq:beta1})--(\ref{eq:beta4}),
and (\ref{eq:charge}) self-consistently at a given baryon density $\rho_{B}$.
Then we can calculate the EOS and the composition of neutron star matter
as well as the effective mass and the Fermi momentum of $\Lambda$ hyperons,
which are crucial in the study of $\Lambda\Lambda$ pairing.


\section{ Gap equation and $\Lambda\Lambda$ interaction}

\label{sec:3}

We study the $^{1}S_{0}$ superfluidity of $\Lambda $ hyperons in neutron
star matter. The crucial quantity in determining the onset of superfluidity
is the energy gap function $\Delta \left( k\right) $, which can be obtained
by solving the gap equation
\begin{equation}
\Delta \left( k\right) =-\frac{1}{4\pi ^{2}}\int k^{\prime 2}dk^{\prime }
\frac{V\left( k,k^{\prime }\right) \Delta \left( k^{\prime }\right) }
{\sqrt{\left[ E\left( k^{\prime }\right) -E\left( k_{F}^{\Lambda }\right)
\right]^{2}+\Delta ^{2}\left( k^{\prime }\right) }},
\label{gapeq}
\end{equation}
where $E\left( k\right) $ is the single-particle energy of $\Lambda$
with momentum $k$. For $\Lambda $ hyperons in neutron star matter,
the single-particle energy in the RMF approach is given by
\begin{equation*}
E\left( k\right) =\sqrt{{k}^{2}+m_{\Lambda }^{\ast 2}}
+g_{\omega\Lambda}\omega +g_{\phi\Lambda}\phi .
\end{equation*}
The effective mass $m_{\Lambda }^{\ast }$ and the Fermi momentum
$k_{F}^{\Lambda }$ are computed self-consistently at a given
baryon density $\rho_{B}$ within the RMF approach.

For the $\Lambda\Lambda$ pairing interaction in the $^{1}S_{0}$
channel, the potential matrix element can be written as
\begin{equation}
V\left( k,k^{\prime }\right) =\left\langle k\right\vert
V_{\Lambda \Lambda}\left( ^{1}S_{0}\right) \left\vert k^{\prime }\right\rangle
=4\pi \int r^{2}dr\text{ }j_{0}(kr)V_{\Lambda \Lambda }
\left( r\right) j_{0}(k^{\prime}r),
\end{equation}
where $j_{0}(kr)=\textrm{sin}(kr)/(kr)$ is the spherical Bessel function
of order zero and $V_{\Lambda \Lambda }\left( r\right) $ is the
$^{1}S_{0}$ $\Lambda\Lambda $ interaction potential in coordinate space.
Because of large uncertainties in the $\Lambda\Lambda$ interaction,
we adopt several $\Lambda\Lambda$ potentials. Most of them are based on
the Nijmegen models and are used in double-$\Lambda$ hypernuclei studies,
which are of the three-range Gaussian form
\begin{equation}
V_{\Lambda \Lambda }(r)=\sum_{i=1}^{3}v_{i}\exp (-r^{2}/\beta_{i}^{2}).
\label{eq:vll}
\end{equation}
The short-range term provides for a strong soft-core repulsion, whereas the
medium-range and long-range terms provide for attraction. The parameters $v_{i}$
and $\beta_{i}$ are taken from Refs.~\cite{Hiy97,NPA02,Hiy02,PRL02}, and we
list them in Table~\ref{tab:1}. The ND1 potential was given in Ref.~\cite{Hiy97}
as an effective soft-core interaction fitted to the Nijmegen model D (ND)
hard-core interaction. Another simulation of the ND interaction, called ND2 in
this article, was obtained in Ref.~\cite{NPA02}. The ESC00, NSC97b, NSC97e,
and NSC97f potentials given in Ref.~\cite{NPA02} were obtained by changing
the strength of the medium-range attractive component of the three-range
Gaussian potential such that they could reproduce the scattering length and
the effective range as close to values by the corresponding Nijmegen models.
It is well known that the Nagara event provides unambiguous identification
of $_{\Lambda \Lambda }^{6}\textrm{He}$ production with a precise
$\Lambda\Lambda$ binding energy value $B_{\Lambda\Lambda }$ and has had a
significant impact on strangeness nuclear physics. The NFs and NSC97s
potentials given in Refs.~\cite{Hiy02,PRL02} were obtained by adjusting
parameters to reproduce the experimental value of
$B_{\Lambda\Lambda}(_{\Lambda\Lambda}^{6}\textrm{He})$ from the Nagara event.
We have also chosen an Urbana-type potential that has
been successfully used to explain the experimental values of hypernuclei.
The Urbana potential could be found in Ref.~\cite{PRC06}.

\begin{figure}[htb]
\includegraphics[bb=20 290 530 775, width=8.6 cm,clip]{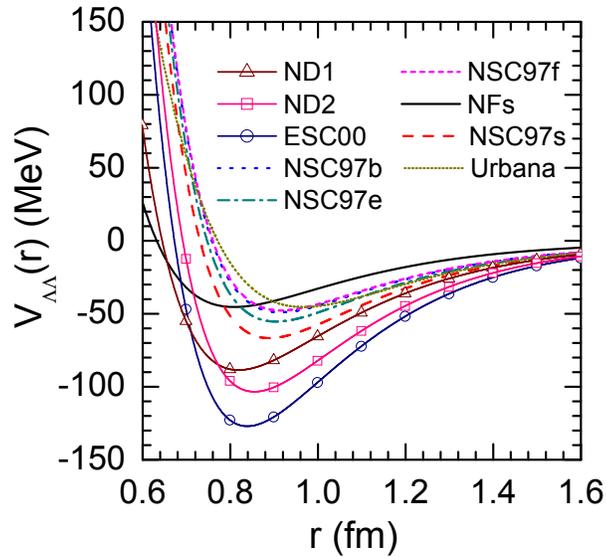}
\caption{(Color online) $^{1}S_{0}$ $\Lambda\Lambda$ interaction potentials
used in this article.}
\label{fig:VR}
\end{figure}

We plot in Fig.~\ref{fig:VR} all $\Lambda\Lambda$ potentials
considered in the present work. The strongest $\Lambda\Lambda$ interaction
is the ESC00 potential, whereas the weakest $\Lambda\Lambda$ interaction
is the NSC97f potential.
We note that the NFs, NSC97s, and Urbana potentials simulate the experimental
value of $B_{\Lambda\Lambda}(_{\Lambda\Lambda}^{6}\textrm{He})$
from the Nagara event.


\section{ Results and discussion}

\label{sec:4}

In this section, we investigate the $^1S_0 $ superfluidity of
$\Lambda$ hyperons in neutron star matter and neutron stars.
We employ the RMF model with the parameter sets TM1 and NL3
to calculate the properties of neutron star matter, which is known
to provide excellent descriptions of the ground states of finite
nuclei, including unstable nuclei.
The meson-hyperon couplings play an important role in determining
the properties of neutron star matter. We use the values constrained
by reasonable hyperon potentials that include the updated information
from recent developments in hypernuclear physics.
As for the $\Lambda\Lambda$ pairing interaction used in the gap equation,
we adopt several $\Lambda\Lambda$ potentials that have been used in
double-$\Lambda$ hypernuclei studies. Some simulate the
experimental value of $B_{\Lambda\Lambda}(_{\Lambda\Lambda}^{6}\textrm{He})$
from the Nagara event. With the effective mass and the Fermi momentum
of $\Lambda$ hyperons obtained in the RMF approach, the gap
equation [Eq.~(\ref{gapeq})] is solved numerically.

\begin{figure}[htb]
\includegraphics[bb=20 290 530 775, width=8.6 cm,clip]{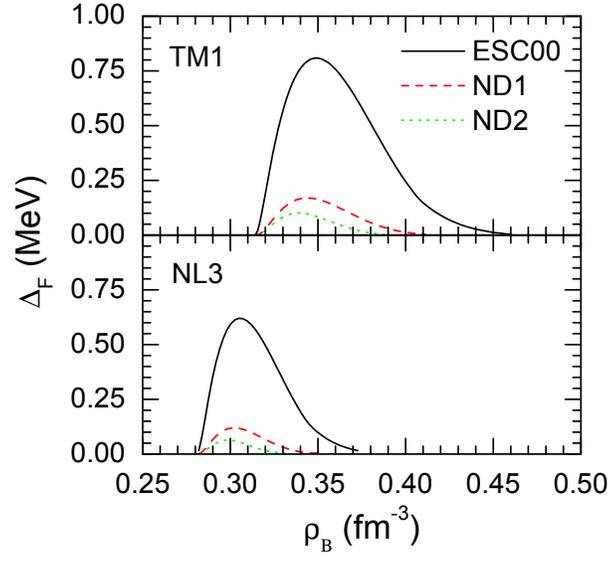}
\caption{(Color online) $^1S_0$ pairing gap of $\Lambda$ hyperons at the
Fermi surface $\Delta_{F}$ as a function of baryon density $\rho_{B}$ in
neutron star matter with the ND1, ND2, and ESC00 potentials:
(top) TM1 and (bottom) NL3.}
\label{fig:RES}
\end{figure}

In Fig.~\ref{fig:RES}, we show the resulting $^1S_0$ pairing gap of
$\Lambda$ hyperons at the Fermi surface, $\Delta_{F}$, as a function of
the baryon density, $\rho_{B}$, in neutron star matter. The results of TM1 and
NL3 are plotted in Fig.~\ref{fig:RES} (top) and Fig.~\ref{fig:RES} (bottom),
respectively. In the case of TM1 (NL3), the threshold density of $\Lambda$
is around $0.31\;\textrm{fm}^{-3}$ ($0.28\;\textrm{fm}^{-3}$),
and $\Lambda$ hyperons form a $^1S_0$ superfluid as soon as they appear
in neutron star matter. With increasing baryon density, $\Delta_{F}$ increases
first, reaching a maximum value at $\rho_{B} \sim 0.34\;\textrm{fm}^{-3}$
($\rho_{B} \sim 0.30\;\textrm{fm}^{-3}$), then decreases and finally vanishes
at $\rho_{B}< 0.46\;\textrm{fm}^{-3}$ ($\rho_{B}< 0.38\;\textrm{fm}^{-3}$)
for the case of TM1 (NL3) with the ESC00 potential.
It is found that the maximal pairing gap is about 0.8 MeV with the
ESC00 potential in the TM1 case. This is because the ESC00 potential
has the strongest attraction among the $\Lambda\Lambda$ interactions used here.
The pairing gaps with the ND1 and ND2 potentials are of the order of
$0.1$--$0.2$ MeV, as shown in Fig.~\ref{fig:RES}. In addition, we find that the
pairing gaps are of the order of $10^{-4}$ MeV (TM1) or absent (NL3)
with the NSC97e, NFs, NSC97s, and Urbana potentials.
The $\Lambda$ pairing does not appear for the NSC97b and NSC97f potentials.
We present in Table~\ref{tab:2} the maximal pairing gap at the Fermi
surface ($\Delta_F^{\textrm{max}}$)
and the corresponding baryon density ($\rho_B$), effective $\Lambda$
mass ($m_\Lambda^\ast$), and Fermi momentum ($k_F^\Lambda$)
using these potentials with the TM1 and NL3 parameter sets.

\begin{figure}[htb]
\includegraphics[bb=20 290 530 775, width=8.6 cm,clip]{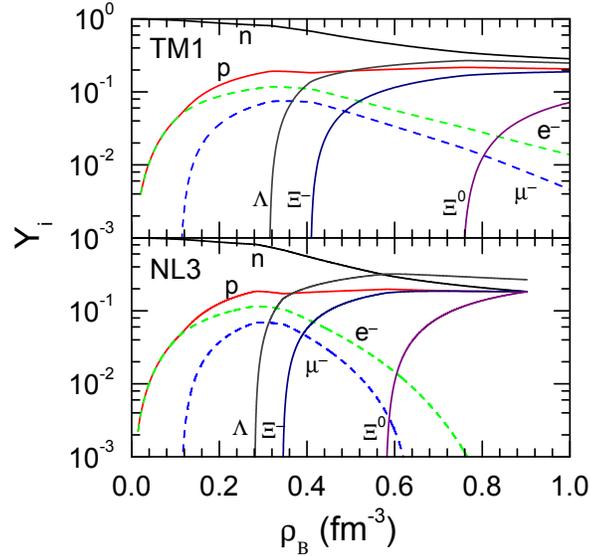}
\caption{(Color online) Particle fraction $Y_{i}=\rho_{i}/\rho_{B}$ as a function
of baryon density $\rho_{B}$: (top) TM1 and (bottom) NL3.}
\label{fig:YI}
\end{figure}

\begin{figure}[htb]
\includegraphics[bb=20 290 530 775, width=8.6 cm,clip]{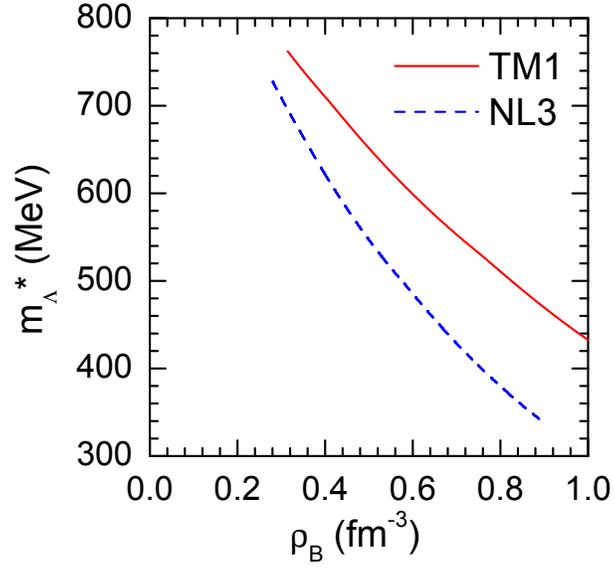}
\caption{(Color online) Effective mass of $\Lambda$ hyperons $m_{\Lambda}^{\ast}$
as a function of baryon density $\rho_{B}$.}
\label{fig:MSTAR}
\end{figure}

The $\Lambda$ pairing gap $\Delta_{F}$ depends not only on the $\Lambda\Lambda$
interaction but also on the properties of $\Lambda$ hyperons in neutron
star matter. In Fig.~\ref{fig:YI}, we show the particle fraction,
$Y_i=\rho_i/\rho_B$, as a function of the baryon density, $\rho_{B}$, using the
RMF model with the TM1 (Fig.~\ref{fig:YI}, top) and NL3 (Fig.~\ref{fig:YI}, bottom)
parameter sets.
It is seen that $\Lambda$ hyperons appear around $0.31\;\textrm{fm}^{-3}$ (TM1)
or $0.28\;\textrm{fm}^{-3}$ (NL3) and then increase rapidly with increasing density.
We note that hyperon threshold densities, fractions, and effective masses are
dependent on the RMF parameters used. This dependence has an effect on
the resulting pairing gap, as shown in Fig.~\ref{fig:RES}.
Our results with the ND1 potential can be compared with those in Table III
of Ref.~\cite{TNYT06}, where the same $\Lambda\Lambda$ interaction
(called the ND-Soft potential) was used. The difference is in the treatment
of neutron star matter, for which they use a nonrelativistic $G$ matrix-based
effective interaction approach, whereas we use the RMF approach.
In our case of TM1 (NL3), the maximal pairing gap at the Fermi surface is
0.17 MeV (0.12 MeV), as given in Table~\ref{tab:2},
where $\rho_B=0.344\;\textrm{fm}^{-3}$ ($\rho_B=0.303\;\textrm{fm}^{-3}$),
$Y_\Lambda=0.039$ ($Y_\Lambda=0.044$), and
$m_{\Lambda}^{\ast}=743$ MeV ($m_{\Lambda}^{\ast}=706$ MeV).
Takatsuka \textit{et al}.~\cite{TNYT06} obtained the maximal pairing gap of 0.34 MeV at
$\rho_B=4.5\rho_{0}$ for the TNI6u EOS. The larger pairing gap at higher
$\rho_B$ given in Ref.~\cite{TNYT06} is because of their smaller $Y_\Lambda$
and larger $m_{\Lambda}^{\ast}$.
As discussed in Refs.~\cite{BB98,TMS03,TNYT06}, the pairing gap is very
sensitive to the effective mass. Generally, a smaller effective mass leads to a
higher single-particle energy and then yields a smaller pairing gap.
In Fig.~\ref{fig:MSTAR}, we show the effective mass of $\Lambda$ hyperons,
$m_{\Lambda}^{\ast}$, as a function of the baryon density, $\rho_{B}$,
using the RMF model with the TM1 (solid line) and NL3 (dashed line) parameter sets.
It is shown that $m_{\Lambda}^{\ast}$ decreases with increasing $\rho_{B}$.
When $\Lambda$ hyperons appear around $0.31\;\textrm{fm}^{-3}$ (TM1)
or $0.28\;\textrm{fm}^{-3}$ (NL3), the effective mass of $\Lambda$ hyperons
is about 762 MeV or 727 MeV. It is found that the effective masses of $\Lambda$
hyperons in the NL3 case are smaller than those of TM1, which leads to a
smaller pairing gap, as shown in Fig.~\ref{fig:RES}.
We note that the effective mass of $\Lambda$ hyperons is mainly determined by
the coupling constants $g_{\sigma\Lambda}$ and $g_{\sigma^{\ast}\Lambda}$.
Here we use the values constrained by reasonable hyperon potentials,
which are suggested by the experimental data of single-$\Lambda$ hypernuclei
and by the Nagara event.

\begin{figure}[htb]
\includegraphics[bb=30 50 530 820, width=8.6 cm,clip]{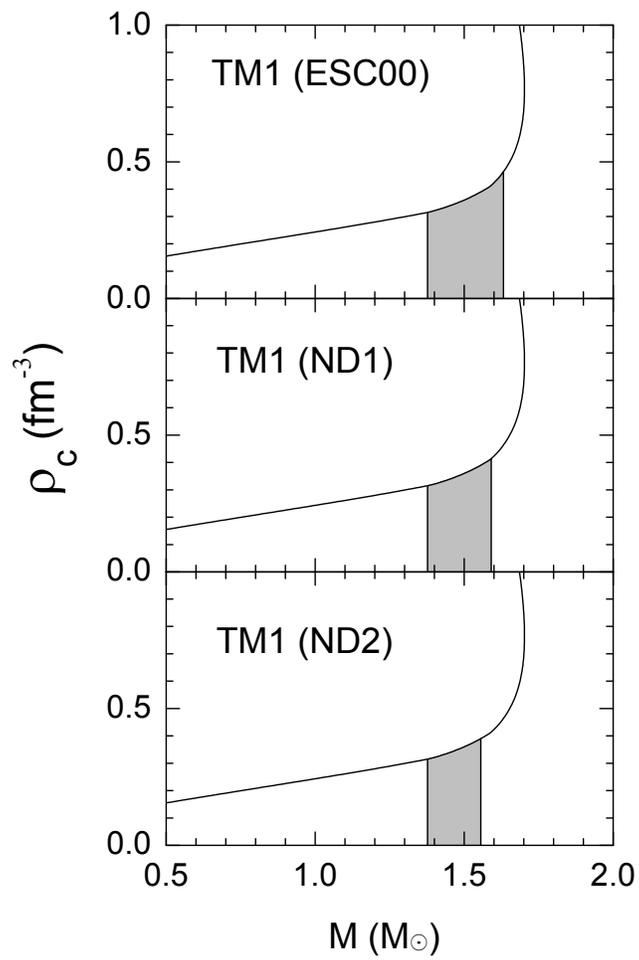}
\caption{Central baryon density $\rho_c$ as a function of
neutron star mass $M$ in the TM1 case.
The region where only superfluid $\Lambda$ hyperons
exist in the core of neutron stars is shaded.}
\label{fig:allT}
\end{figure}

\begin{figure}[htb]
\includegraphics[bb=30 50 530 820, width=8.6 cm,clip]{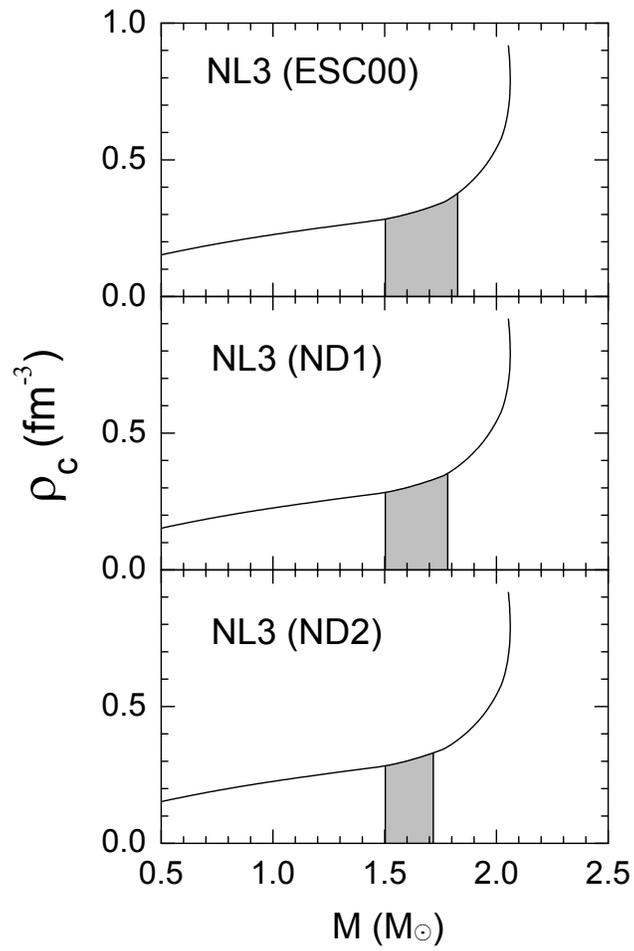}
\caption{Same as Fig.~\ref{fig:allT}, but for the NL3 case.}
\label{fig:allN}
\end{figure}

To examine whether the $^1S_0 $ superfluidity of $\Lambda$ hyperons exists
in neutron stars, we solve the Tolman-Oppenheimer-Volkoff (TOV) equation
with the EOS of the RMF theory over a wide density range.
For the nonuniform matter at low density, which exists in the inner and
outer crusts of neutron stars, we adopt a relativistic EOS based on the
RMF theory with a local density approximation~\cite{Shen02,Shen98}.
The nonuniform matter is modeled to be composed of a lattice of
spherical nuclei immersed in an electron gas with or without
free neutrons dripping out of nuclei. The low-density EOS is
matched to the EOS of uniform matter at the density
where they have equal pressure. The neutron star properties
are mainly determined by the EOS at high density.
Using the EOS described in Sec.~\ref{sec:2}, we calculate the neutron
star properties and find that the maximum mass of neutron stars
is about 1.70 $M_{\odot}$ (2.06 $M_{\odot}$)
with the TM1 (NL3) parameter set.
According to the compilation of measured neutron star masses~\cite{PR07,NS09},
some massive neutron stars were reported to be observed recently.
However, the uncertainties in these mass measurements are rather large,
and the mass of PSR J0751$+$1807 was corrected from
$(2.1 \pm 0.2)$ $M_{\odot}$ to $(1.26 \pm 0.14)$ $M_{\odot}$~\cite{NS08}.
We note that the EOS used here could not be ruled out by current observations.
In Figs.~\ref{fig:allT} and~\ref{fig:allN},
we show the central baryon density as a function of the neutron star mass.
We find that whether the $^1S_0 $ superfluidity of $\Lambda$ hyperons
exists in the core of neutron stars depends on the $\Lambda\Lambda$
interaction used. With weaker $\Lambda\Lambda$ interactions, such as
NSC97b and NSC97f, the $\Lambda$ superfluidity does not appear inside neutron
stars. For the NSC97e, NFs, NSC97s, and Urbana interactions,
although we obtain the pairing gaps of the order of $10^{-4}$ MeV in
the TM1 case, it is unlikely that $\Lambda$ superfluidity can exist
in observed neutron stars because of its low superfluid critical temperature
$T_c \simeq 0.57 \Delta_{F} / \kappa_B \sim 10^6$ K~\cite{TNYT06,Yakovlev01}.
With stronger $\Lambda\Lambda$ interactions, such as ESC00, ND1, and ND2,
the $^1S_0 $ superfluidity of $\Lambda$ hyperons may exist
in massive neutron stars, as shown in Figs.~\ref{fig:allT} and~\ref{fig:allN}.
In the case of TM1 (NL3) with the ESC00 potential,
$\Lambda$ hyperons do not appear in neutron stars
with $M < 1.37 \ M_{\odot}$ ($M < 1.50 \ M_{\odot}$).
For neutron stars with $1.37 \ M_{\odot} < M < 1.63 \ M_{\odot}$
($1.50 \ M_{\odot} < M < 1.82 \ M_{\odot}$), $\Lambda$ hyperons
in the core of neutron stars form a $^1S_0$ superfluid.
However, when $M > 1.63 \ M_{\odot}$ ($M > 1.82 \ M_{\odot}$),
not only superfluid $\Lambda$ but also normal (nonsuperfluid)
$\Lambda$ can exist in the core of neutron stars
because the central baryon density exceeds the upper limit
of the range where $\Lambda$ superfluidity exists.
The presence of nonsuperfluid $\Lambda$ hyperons in the core of
massive stars would lead to a more rapid cooling than the case with
only superfluid $\Lambda$ hyperons.
The mass region, where only superfluid $\Lambda$ hyperons
exist in the core of neutron stars, is shaded
in Figs.~\ref{fig:allT} and~\ref{fig:allN}.
It is shown that the region with the ESC00 potential is the widest
among all cases in these figures. This is because the ESC00 potential
has the strongest attraction, and its pairing gap covers the widest
density range, as shown in Fig.~\ref{fig:RES}.
We note that this region depends both on the $\Lambda\Lambda$
interaction and on the EOS of neutron star matter.


\section{Summary}

\label{sec:5}

We have studied the $^1S_0$ superfluidity of $\Lambda$ hyperons
in neutron star matter and neutron stars. In this article,
we employ the RMF model with the parameter sets TM1 and NL3
to calculate the properties of neutron star matter,
which is composed of a chemically equilibrated and
charge-neutral mixture of nucleons, hyperons, and leptons.
The RMF theory has been successfully and
widely used for the description of nuclear matter and finite nuclei,
including unstable nuclei. In the RMF approach, baryons interact
through the exchange of scalar and vector mesons.
The baryons considered in this article are nucleons ($p$ and $n$)
and hyperons ($\Lambda$, $\Sigma$, and $\Xi$).
The exchanged mesons include isoscalar scalar and vector mesons
($\sigma$ and $\omega$), an isovector vector meson ($\rho$), and two
additional hidden-strangeness mesons ($\sigma^{\ast }$ and $\phi$).
It is well known that the meson-hyperon couplings play an important
role in determining the properties of neutron star matter.
We have used the couplings constrained by reasonable hyperon
potentials that include the updated information from recent
developments in hypernuclear physics.
To examine the $^1S_0$ pairing of $\Lambda$ hyperons, we have adopted
several $\Lambda\Lambda$ potentials.  Most are based on the
Nijmegen models and have been used in double-$\Lambda$ hypernuclei studies.
NFs, NSC97s, and Urbana potentials have simulated the experimental
value of $B_{\Lambda\Lambda}(_{\Lambda\Lambda}^{6}\textrm{He})$
from the Nagara event.

We have calculated the $^1S_0$ pairing gap of $\Lambda$ hyperons at the
Fermi surface, $\Delta_{F}$, using the $\Lambda\Lambda$ potentials
adopted in this article. It is found that $\Delta_{F}$ depends
both on the $\Lambda\Lambda$ interaction and on the treatment
of neutron star matter. The maximal $\Delta_{F}$ obtained in
the present calculation is about 0.8 MeV with the ESC00 potential in the TM1 case.
This is because the ESC00 potential has the strongest attraction
among the $\Lambda\Lambda$ interactions used in this article.
The ND1 and ND2 potentials yield somewhat smaller $\Delta_{F}$ of the order
of $0.1$--$0.2$ MeV. For the NSC97e, NFs, NSC97s, and Urbana potentials,
the values of $\Delta_{F}$ are of the order of $10^{-4}$ MeV (TM1) or absent (NL3).
The $\Lambda$ pairing does not appear for the NSC97b and NSC97f potentials.
The difference in these results reflects the dependence of $\Delta_{F}$
on the $\Lambda\Lambda$ interaction.
On the other hand, the magnitude and the threshold density of $\Delta_{F}$
are also dependent on properties of neutron star matter, especially
on the effective mass and particle fraction of $\Lambda$ hyperons.
In the case of TM1 (NL3) with the ESC00 potential,
the threshold density of $\Delta_{F}$ is around
$0.31\;\textrm{fm}^{-3}$ ($0.28\;\textrm{fm}^{-3}$),
reaches a maximum value at $\rho_{B} \sim 0.34\;\textrm{fm}^{-3}$
($\rho_{B} \sim 0.30\;\textrm{fm}^{-3}$), and finally vanishes at
$\rho_{B}< 0.46\;\textrm{fm}^{-3}$ ($\rho_{B}< 0.38\;\textrm{fm}^{-3}$).
By solving the TOV equation, we have calculated neutron star properties
and found that whether the $^1S_0 $ superfluidity of $\Lambda$ hyperons
exists in the core of neutron stars mainly depends on the $\Lambda\Lambda$
interaction used. With stronger $\Lambda\Lambda$ interactions, such as
ESC00, ND1, and ND2, the $\Lambda$ superfluidity may exist in massive neutron
stars. It is unlikely that $\Lambda$ superfluidity can exist
in neutron stars with the NFs, NSC97s, and Urbana interactions,
which have simulated the experimental value of
$B_{\Lambda\Lambda}(_{\Lambda\Lambda}^{6}\textrm{He})$
from the Nagara event.

In this article, we have considered the updated information from recent
developments in hypernuclear physics and used the weak attractive
$\Lambda\Lambda$ interactions suggested by the Nagara event.
However, there are still large uncertainties in the hyperon-hyperon
interaction and the EOS of neutron star matter.
A more precise study of the $\Lambda$ pairing in neutron stars
requires further development in hypernuclear physics.


\section*{ACKNOWLEDGMENT}

This work was supported in part by the National Natural Science
Foundation of China (Grant No. 10675064).


\newpage

\begin{table}[!h]
\caption{Parameters of $^{1}S_{0}$ $\Lambda\Lambda$ interaction
defined in Eq.~(\ref{eq:vll}), taken from
Refs.~\cite{Hiy97,Hiy02,NPA02,PRL02}. The size parameters are the same
for all cases, which are
$\beta_{1}=1.342$ fm, $\beta_{2}=0.777$ fm, and $\beta_{3}=0.350$ fm.
The strength parameters are in MeV.}
\label{tab:1}
\begin{center}
\begin{tabular}{llcccccccccccccccccccccc}
\hline\hline
                      & & & $v_{1}$ & & & $v_{2}$ & & & $v_{3}$\\
\hline
               ND1    & & & -21.92 & & & -283.5 & & & 4745\\
               ND2    & & & -21.49 & & & -379.1 & & & 9324\\
               ESC00  & & & -21.49 & & & -456.6 & & & 9324\\
               NSC97b & & & -21.49 & & & -182.1 & & & 9324\\
               NSC97e & & & -21.49 & & & -207.1 & & & 9324\\
               NSC97f & & & -21.49 & & & -177.1 & & & 9324\\
               NFs    & & & -10.96 & & & -141.8 & & & 2137\\
               NSC97s & & & -21.49 & & & -250.1 & & & 9324\\
\hline\hline
\end{tabular}
\end{center}
\end{table}

\begin{table}[!h]
\caption{Maximal pairing gap at the Fermi surface $\Delta_F^{\textrm{max}}$
obtained with several $\Lambda\Lambda$ potentials;
$\rho_B$ is the total baryon density of neutron star matter
where $\Delta_F^{\textrm{max}}$ is obtained, and $k_F^\Lambda$ and $m_\Lambda^\ast$
are the corresponding Fermi momentum and effective mass of $\Lambda$
hyperons, respectively.}
\label{tab:2}
\begin{center}
\begin{tabular}{lcccccccccc}
\hline\hline
& & TM1 & & & & & & NL3\\
\hline
& $\rho_B$ & $k_F^\Lambda$    & $m_\Lambda^\ast$ & $\Delta_F^{\textrm{max}}$& & & $\rho_B$ & $k_F^\Lambda$    & $m_\Lambda^\ast$ & $\Delta_F^{\textrm{max}}$\\
             & ($\textrm{fm}^{-3}$) &($\textrm{fm}^{-1}$)& (MeV) & (MeV)  & & & ($\textrm{fm}^{-3}$) &($\textrm{fm}^{-1}$)& (MeV) & (MeV)  \\
\hline
      ND1    & 0.344 & 0.738 & 743 & 0.17 & & & 0.303 & 0.731 & 706 & 0.12\\
      ND2    & 0.339 & 0.681 & 747 & 0.10 & & & 0.298 & 0.664 & 711 & 0.06\\
      ESC00  & 0.349 & 0.789 & 740 & 0.81 & & & 0.305 & 0.762 & 704 & 0.62\\
      NSC97b & - & - & - & - & & & - & - & - & -\\
      NSC97e & 0.329 & 0.548 & 753 & $1.2\times 10^{-4}$ & & & - & - & - & -\\
      NSC97f & - & - & - & - & & & - & - & - & -\\
      NFs    & 0.329 & 0.548 & 753 & $5.4\times 10^{-4}$ & & & - & - & - & -\\
      NSC97s & 0.329 & 0.548 & 753 & $4.0\times 10^{-4}$ & & & - & - & - & -\\
      Urbana & 0.329 & 0.548 & 753 & $5.5\times 10^{-4}$ & & & - & - & - & -\\
\hline\hline
\end{tabular}
\end{center}
\end{table}

\newpage

\end{document}